\documentstyle[12pt]{article}

\newfont{\bbd}{msbm10 scaled\magstep1}

\begin{document}
\thispagestyle{empty}

\newcommand{\p}[1]{(\ref{#1})}
\newcommand{\be}{\begin{equation}}
\newcommand{\ee}{\end{equation}}
\newcommand{\sect}[1]{\setcounter{equation}{0}\section{#1}}

\newcommand{\vs}[1]{\rule[- #1 mm]{0mm}{#1 mm}}
\newcommand{\hs}[1]{\hspace{#1mm}}
\newcommand{\mb}[1]{\hs{5}\mbox{#1}\hs{5}}
\newcommand{\Db}{{\overline D}}
\newcommand{\bea}{\begin{eqnarray}}
\newcommand{\eea}{\end{eqnarray}}
\newcommand{\wt}[1]{\widetilde{#1}}
\newcommand{\und}[1]{\underline{#1}}
\newcommand{\ov}[1]{\overline{#1}}
\newcommand{\sm}[2]{\frac{\mbox{\footnotesize #1}\vs{-2}}
		   {\vs{-2}\mbox{\footnotesize #2}}}
\newcommand{\prt}{\partial}
\newcommand{\eps}{\epsilon}

\newcommand{\R}{\mbox{\rule{0.2mm}{2.8mm}\hspace{-1.5mm} R}}
\newcommand{\Z}{Z\hspace{-2mm}Z}

\newcommand{\cd}{{\cal D}}
\newcommand{\cg}{{\cal G}}
\newcommand{\ck}{{\cal K}}
\newcommand{\cw}{{\cal W}}

\newcommand{\vj}{\vec{J}}
\newcommand{\vl}{\vec{\lambda}}
\newcommand{\vz}{\vec{\sigma}}
\newcommand{\vt}{\vec{\tau}}
\newcommand{\vw}{\vec{W}}
\newcommand{\poiss}{\stackrel{\otimes}{,}}

\def\l#1#2{\raisebox{.2ex}{$\displaystyle
  \mathop{#1}^{{\scriptstyle #2}\rightarrow}$}}
\def\r#1#2{\raisebox{.2ex}{$\displaystyle
 \mathop{#1}^{\leftarrow {\scriptstyle #2}}$}}

\renewcommand{\thefootnote}{\fnsymbol{footnote}}
\newpage
\setcounter{page}{0}
\pagestyle{empty}
\begin{flushright}
{January 2002}\\
{nlin.SI/0201061}\\
\end{flushright}
\vs{8}
\begin{center}
{\LARGE {\bf Bi-Hamiltonian structure of the}}\\[0.6cm]
{\LARGE {\bf N=2 supersymmetric ${\alpha}=1$ KdV hierarchy}}\\[1cm]

\vs{8}

{\large P.H.M. Kersten$^{(a)}$ and A.S. Sorin$^{(b)}$}
{}~\\
\quad \\
{\em ~$~^{(a)}$ University of Twente, Faculty of Mathematical Sciences, \\
P.O.Box 217,7500 AE Enschede, The Netherlands \\
E-Mail: kersten@math.utwente.nl}\\
{}~\\
{\em {~$~^{(b)}$ Bogoliubov Laboratory of Theoretical Physics, \\
Joint Institute for Nuclear Research (JINR),}}\\
{\em 141980 Dubna, Moscow Region, Russia \\
E-Mail: sorin@thsun1.jinr.ru}~\quad\\

\end{center}
\vs{8}

\centerline{ {\bf Abstract}}
The $N=2$ supersymmetric ${\alpha}=1$ KdV hierarchy in $N=2$ superspace
is considered and its rich symmetry structure is uncovered. New
nonpolynomial and nonlocal, bosonic and fermionic symmetries and
Hamiltonians, bi-Hamiltonian structure as well as a recursion operator
connecting all symmetries and Hamiltonian structures of the $N=2$
${\alpha}=1$ KdV hierarchy are constructed in explicit form. It is 
observed that the algebra of symmetries of the $N=2$ supersymmetric
${\alpha}=1$ KdV hierarchy possesses two different subalgebras of $N=2$
supersymmetry.  

{}~

{}~

{}~


Recursion operator; 
Bi-Hamiltonian structure

\newpage

\pagestyle{plain}
\renewcommand{\thefootnote}{\arabic{footnote}}
\setcounter{footnote}{0}

\section{\bf Introduction}
The $N=2$ supersymmetric ${\alpha}=1$ KdV equation was originally 
introduced in \cite{lm} as a Hamiltonian equation with the $N=2$
superconformal algebra as a second Hamiltonian structure, and 
its integrability was conjectured there due to the existence of 
a few additional nontrivial bosonic Hamiltonians. Then its Lax--pair
representation has indeed been constructed in \cite{pop1}, and it 
allowed an algoritmic reconstruction of the whole tower of highest 
{\it commutative} bosonic flows and their Hamiltonians belonging to the
$N=2$ supersymmetric ${\alpha}=1$ KdV hierarchy. 

Actually, besides the $N=2$ ${\alpha}=1$ KdV equation
there are another two inequivalent $N=2$ supersymmetric 
Hamiltonian equations with the $N=2$ superconformal algebra as a
second Hamiltonian structure (the $N=2$ ${\alpha}=-2$ and
${\alpha}=4$ KdV equations \cite{mat,lm}), but the $N=2$ ${\alpha}=1$
KdV equation is rather exceptional \cite{bks}. Despite knowledge of its
Lax--pair description, there remains a lot of longstanding, unsolved
problems which resolution would be quite important for a deeper
understanding and more detailed description of the $N=2$ ${\alpha}=1$ KdV
hierarchy. Thus,
since the time when the $N=2$ ${\alpha}=1$ KdV equation was proposed, much
efforts were made to construct a tower of its {\it noncommutative} bosonic
and fermionic, local and nonlocal symmetries and Hamiltonians,
bi-Hamiltonian structure as well as recursion operator (see, e.g.
discussions in \cite{dm,m} and references therein). Though these
rather complicated problems, solved for the case of
the $N=2$ ${\alpha}=-2$ and ${\alpha}=4$ KdV hierarchies,
still wait their complete resolution for the $N=2$ ${\alpha}=1$ KdV 
hierarchy, a considerable progress towards their solution arose quite
recently. Thus, the {\it puzzle} \cite{dm,m}, related to the 
{\it "nonexistence"} 
of higher fermionic flows of the $N=2$ ${\alpha}=1$ KdV hierarchy,
was partly resolved in \cite{ker,ker1} by explicit constructing a few
bosonic and fermionic {\it nonlocal} and {\it nonpolynomial} flows
and Hamiltonains, then their $N=2$ superfield structure and origin were
uncovered in \cite{ks}. A new property, crucial for the existence of these
flows and Hamiltonians, making them distinguished compared to all
flows and Hamiltonians of other supersymmetric hierarchies 
constructed before, is their {\it nonpolynomiality}.
A new approach to a recursion operator treating it as a form--valued
vector field which satisfies a generalized symmetry equation related to
a given equation was developed in \cite{kr1, kerkr2}. Using this approach
the recursion operator of the bosonic limit of the N=2
${\alpha}=1$ KdV hierarchy was derived in \cite{ker2}, and its structure,
underlining relevance of these Hamiltonians in the bosonic limit,  
gives a hint towards its supersymmetric generalization.

The present Letter addresses the above--mentioned problems. 
We demonstrate that the existence and knowledge of the bosonic and
fermionic, nonlocal and nonpolynomial Hamiltonians of the $N=2$
${\alpha}=1$ KdV hierarchy is indeed a crucial, key point in constructing
the recursion operator connecting all its symmetries and Hamiltonian
structures. 

The Letter is organized as follows. In Section 2 we present a short
summary of the main known facts concerning the $N=2$ ${\alpha}=1$ KdV
hierarchy. In Section 3 we describe a general algorithm of constructing
recursion operators of integrable systems which we follow
in the next sections. In Section 4 we construct a minimal, basic set of
fermionic and bosonic flows as well as their Hamiltonians which is then
used in Section 5 to derive the recursion operator and
bi-Hamiltonian structure of the $N=2$ ${\alpha}=1$ KdV hierarchy.
We also demonstrate that the algebra of symmetries of the $N=2$
${\alpha}=1$ KdV hierarchy possesses two different subalgebras of $N=2$
supersymmetry. In Section 6 we summarize our results.

\section{\bf The N=2 supersymmetric ${\alpha}=1$ KdV hierarchy}

The $N=2$ supersymmetric ${\alpha}=1$ KdV equation \cite{lm}
\begin{eqnarray}
{\textstyle{\partial\over\partial t_3}}{J} =
[J~'' + 3 J[D,{\overline D}~] J + J^3]~'
\label{flows}
\end{eqnarray}
is the first nontrivial representative of the infinite tower of the
commutative flows ${\textstyle{\partial\over\partial t_{2p+1}}}$ 
\begin{eqnarray}
[{\textstyle{\partial\over\partial t_{2m+1}}},
{\textstyle{\partial\over\partial t_{2n+1}}}]=0
\label{alg}
\end{eqnarray}
belonging to the $N=2$ supersymmetric ${\alpha}=1$ KdV hierarchy.  
The latter can be defined via the Lax--pair representation \cite{pop1} 
\begin{eqnarray}
{\textstyle{\partial\over\partial t_{2p+1}}}L = 
[ (L^{2p+1})_{\geq 0}, L ],\quad
L =\partial + [D,{\overline D}]{\partial}^{-1}J,\quad p \in \hbox{\bbd N}   
\label{lax}
\end{eqnarray}
where the subscript $\geq 0$ denotes the sum of purely differential and
constant parts of the operator $L^{2p+1}$,
$J\equiv J(Z)$ is a superfield of inverse length dimension $[J]=1$
in the $N=2$ superspace with a
coordinate $Z=(z,\theta,\overline\theta)$,
$'$ denotes the derivative with respect to $z$
and $D,{\overline D}$ are the
fermionic covariant derivatives of $N=2$ supersymmetry 
\begin{eqnarray}
D=\frac{\partial}{\partial\theta}
 -\frac{1}{2}\overline\theta\frac{\partial}{\partial z}, \quad
{\overline D}=\frac{\partial}{\partial\overline\theta}
 -\frac{1}{2}\theta\frac{\partial}{\partial z}, \quad
D^{2}={\overline D}^{2}=0, \quad
\left\{ D,{\overline D} \right\}= -\frac{\partial}{\partial z}
\equiv -{\partial}.
\label{DD}
\end{eqnarray}

The flows \p{lax} are Hamiltonian
\begin{eqnarray}
{\textstyle{\partial\over\partial t_{2p+1}}}J = \{J, H_{2p+1}\}_2
\equiv J_2 \frac{\delta}{\delta J} H_{2p+1} 
\label{hamstr}
\end{eqnarray}
with the Hamiltonians  
\begin{eqnarray}
H_{2p+1} =\int d Z ~{\cal H}_{2p+1}, \quad {\cal H}_{2p+1}=res(L^{2p+1})
\label{bosham}
\end{eqnarray}
and the Poisson brackets 
\begin{eqnarray}
\{J(Z_1),J(Z_2)\}_2 =J_2(Z_1) \delta^{N=2}(Z_1-Z_2),\quad 
J_2\equiv \frac{1}{2}[D, \overline D ~]\partial +\overline DJD 
+DJ\overline D + \partial J +J\partial
\label{hamstr2}
\end{eqnarray}
forming the $N=2$ superconformal algebra which is the second Hamiltonian
structure of the $N=2$ ${\alpha}=1$ KdV hierarchy \cite{lm}. 
Here, $dZ \equiv dz d \theta d \overline\theta$, 
$\int dZ (...) \equiv \int^{+\infty}_{-\infty} dz D {\overline D}
(...) \Big\vert_{\theta = \overline \theta =0}$, $res(L^p)$ is the
$N=2$ supersymmetric residue, i.e. the coefficient at 
$[D,\overline D]{\partial}^{-1}$, and $\delta^{N=2}(Z)\equiv \theta
\overline \theta \delta(z)$ is the delta function in $N=2$ superspace.

The flows \p{lax} admit also the following useful representation \cite{ks}
\begin{eqnarray}
{\textstyle{\partial\over\partial t_{2p+1}}}J = {\cal H}^{~'}_{2p+1}
\label{bosflows}
\end{eqnarray}
in terms of Hamiltonian densities ${\cal H}_{2p+1}$ \p{bosham} and
possess the complex conjugation
\begin{eqnarray}
(t_{2p+1},z,{\theta}, {\overline {\theta}},J)^{\star}=
(-t_{2p+1},-z,{\overline {\theta}},{\theta},J).
\label{conjstar}
\end{eqnarray}

A few first representatives of the Hamiltonians $H_{2p+1}$ \p{bosham}
and flows ${\textstyle{\partial\over\partial t_{2p+1}}}$ \p{bosflows} are
\begin{eqnarray}
H_{1} =\int d Z ~J, \quad  H_3 = \int dZ~(3J[D,{\overline D}]J+J^3)
\label{bosham03}
\end{eqnarray}
and
\begin{eqnarray}
{\textstyle{\partial\over\partial t_1}}{J} = J~' 
\label{flows1}
\end{eqnarray}
and the third flow ${\textstyle{\partial\over\partial t_3}}{J}$ which
reproduces the $N=2$ supersymmetric ${\alpha}=1$ KdV
equation \p{flows}.

\section{\bf General algorithm of constructing recursion operators}

Let us present a short summary of the general scheme of constructing
recursion operators of integrable systems, considered 
in \cite{ker} (for more details, see
\cite{ker} and references therein), which we adapt and develop
to the problems under consideration and use in next sections.

In effect, in the theory of deformations of the equation structure a
classical recursion operator is a form--valued vector field 
which is a nontrivial infinitesimal deformation of
the equation structure. For reasons of simplicity and in order
to make it more accessible to the physics oriented reader, in the
following the form--valuedness will be transformed to Fr\'echet
derivatives of the associated variables and quantities.
 
Our starting point is the one-dimensional system of evolution 
scale--invariant\footnote{This means that all parameters, involved in the
equation, are dimensionless.} equations
(with the evolution time $t$) for the set of superfields 
$u \equiv u_i(Z)$ $(i=1,...,n)$ 
\begin{eqnarray}
{\textstyle{\partial\over\partial t_{}}}u-\Phi=0,
\label{phi}
\end{eqnarray}
their integrals of motion
\begin{eqnarray}
{\cal G}_{\alpha}=\int^{+\infty}_{-\infty} dz G_{\alpha}, \quad
{\textstyle{\partial\over\partial t_{}}}G_{\alpha}= {F_{\alpha}}~' , \quad
{\alpha}=0,1,...
\label{hi}
\end{eqnarray}
and the corresponding symmetry equation 
\begin{eqnarray}
\Bigl({\textstyle{\partial\over\partial t_{}}}-\phi\Bigr)Y_{\tau}=0,\quad 
Y_{\tau} := {\textstyle{\partial\over\partial {\tau}_{}}}u
\label{symeq}
\end{eqnarray}
which is derived by differentiation of the system \p{phi} over
an arbitrary independent time\footnote{We would like to emphasize that 
supersymmetric systems in general admit both bosonic or fermionic
evolution times. Here, for definiteness we discuss only the former case,
its generalization to the latter case is rather straightforward.}
${\tau}$. Here, $\Phi\equiv \Phi(\{u\})$ is a local, while  
$G_i\equiv G_i(\{u\})$, $F_i\equiv F_i(\{u\})$ as well as 
$Y_{\tau}\equiv Y_{\tau}(\{u\})$ in general are nonlocal functionals 
of \{${\partial}^{n}u,~ D{\partial}^{n}u,~ 
{\overline D}{\partial}^{n}u, [D,{\overline D}]{\partial}^{n}u,
~n \in \hbox{\bbd N}\}$,
and the operator $\phi$ is the corresponding Fr\'echet
derivative of the functional $\Phi$. 

The symmetry equation \p{symeq} represents a complicated 
functional equation, and its general solution is not known.
Its particular solutions are symmetries of the
system \p{phi} we started with, i.e. 
$[{\textstyle{\partial\over\partial t_{}}},
{\textstyle{\partial\over\partial {\tau}_{}}}]=0$ by construction.
For a more complete understanding of the original system, its hierarchy
structure and solutions (tau function) it seems necessary to know as many
solutions of its symmetry equation \p{symeq} as possible. 

It turns out that there exist subsets of the whole set of solutions of the
symmetry equation \p{symeq} for which two different representatives
$Y_{\tau}$ and $Y_{\widetilde {\tau}} $ can consistently be related as
\begin{eqnarray} 
Y_{\widetilde {\tau}} :=
{\textstyle{\partial\over\partial {\widetilde {\tau}}_{}}}u=
P{\textstyle{\partial\over\partial {\tau}_{}}}u +\sum_{\alpha} P_{\alpha}
{\textstyle{\partial\over\partial {\tau}_{}}} {\partial}^{-1}G_{\alpha}
\equiv \Bigl(P+\sum_{{\alpha}} P_{\alpha} {\partial}^{-1}
g_{\alpha}\Bigr)Y_{\tau}:= R Y_{\tau} 
\label{recrel1} 
\end{eqnarray} 
where the operator $g_{\alpha}$ is the Fr\'echet derivative of the
functional $G_{\alpha}$, $P\equiv P(\{u,{\partial}^{-1}G,{\partial},
D,{\overline D}\})$  is a general purely differential operator over the
derivatives $\{{\partial},~ D,~ {\overline D}\}$ which
coefficient--functions are
scale--homogeneous polynomials over 
$\{u,{\partial}^{-1}G\}$ and their 
\{${\partial}^{n},~ D{\partial}^{n},~$
${\overline D}{\partial}^{n},$ $[D,{\overline D}]
{\partial}^{n}\}$--derivatives
and $P_{\alpha}\equiv P_{\alpha}(\{u,
{\partial}^{-1}G\})$ is a scale--homogeneous polynomial functional over
its two arguments and their \{${\partial}^{n},~ D{\partial}^{n},~$
${\overline D}{\partial}^{n},~ [D,{\overline D}]
{\partial}^{n}\}$--derivatives 
obeying the so called
deformation equation\footnote{Here, the brackets $(...)_0$ mean that
the relevant operators act only on the superfields inside the brackets.}
\begin{eqnarray} \Bigl(({\textstyle{\partial\over\partial {t}_{}}}P)
 -\phi P\Bigr){\textstyle{\partial\over\partial {\tau}_{}}}u
+P{\textstyle{\partial\over\partial {\tau}_{}}}\Phi +\sum_{{\alpha}}
\Bigl\{\Bigl((p_{\alpha} \Phi)_0 -\phi P_{\alpha} \Bigr) 
{\textstyle{\partial\over\partial {\tau}_{}}} {\partial}^{-1}G_{\alpha}+
P_{\alpha} {\textstyle{\partial\over\partial {\tau}_{}}}
F_{\alpha}\Bigr\}=0 \label{maineq} \end{eqnarray} which is the consistency
condition resulting from the requirement that both $Y_{\tau}$ and
$Y_{\widetilde {\tau}}$ have to satisfy the symmetry equation \p{symeq}.
Here, the operator $p_{\alpha}$ is the Fr\'echet derivative of the
functional $P_{\alpha}$. By definition, the operator $R$, defined in eq.
\p{recrel1}, which has a minimal inverse length dimension, is the
recursion operator of the hierarchy of symmetries of the equation \p{phi}.
Scale dimensions of the quantities $P$ and $P_{\alpha}$ depend on the
dimension of the recursion operator (see. eqs. \p{dimP}). Extracting the
coefficients at linear--independent functionals 
${\textstyle{\partial\over\partial {\tau}_{}}}u$,
${\textstyle{\partial\over\partial {\tau}_{}}}{\partial}^{-1}G_{\alpha}$
and their \{${\partial}^{n},~
D{\partial}^{n},~ {\overline D}{\partial}^{n}, [D,{\overline
D}]{\partial}^{n}u\}$--derivatives in eq. \p{maineq} and equating them
to zero one can derive a complete set of self--consistent equations
for coefficient--functions of the operator $P$ and polynomials
$P_{\alpha}$ which solutions specify the recursion operator R \p{recrel1}. 

A few important remarks are in order. 

First, coefficient--functions of the operator $P$ and polynomial
$P_{\alpha}$ can in
general be non--polynomial functions of the dimensionless quantity
${\partial}^{-1}G_0$ ($[{\partial}^{-1}G_0]=0$) (if any).  

Second, one can see from a simple dimensional consideration that the sum
over $\alpha$ in eqs. (\ref{recrel1}--\ref{maineq}) is usually saturated
by a finite number of terms due to pure dimensional restrictions. Indeed,
the inverse length dimensions of the quantities entering into eqs.
(\ref{recrel1}--\ref{maineq}) are related and bounded as $[P_{\alpha}]
+[{\cal G}_{\alpha}] -[u]= [R] > 0$, $[P_{\alpha}] \geq 0$, 
$[{\cal G}_{\alpha}]\geq 0$, so the inverse length dimension 
$[{\cal G}_{\alpha}]$
of the integrals ${\cal G}_{\alpha}$ \p{hi}, contributing the sum over
${\alpha}$, is bounded both from below and above as 
\begin{eqnarray} 
0\leq [{\cal G}_{\alpha}]\leq [R]+[u], 
\label{uneq} 
\end{eqnarray} 
but in general there exists only a finite number of integrals 
which inverse length dimensions belong to a finite
interval $\Bigl[0,[R]+[u]\Bigr]$.  

Third, a simple inspection of eq. \p{maineq} shows that the functional
$P_{\alpha}$, which is a factor of the ${\tau}$--derivative of 
the lowest dimension superfield component of the integral\footnote{An
$N=2$ superfield
integral of the form \p{hi} has four independent superfield components in
general.} of highest inverse length dimension entering into eq.
\p{recrel1}, satisfies the symmetry equation \p{symeq}, so this $P_{\alpha}$ 
is a symmetry of eq. \p{phi}.

Summarizing the above--described algorithm, it consists of a few steps.
Thus, at the first step, as an input it is necessary to define the inverse
length dimension $[R]$ of the recursion operator, then to construct a
complete set of integrals ${\cal G}_{\alpha}$ \p{hi}
for eq. \p{phi} which inverse length dimensions
$[{\cal G}_{\alpha}]$ satisfy the inequality \p{uneq}.
The next step is a rather straightforward technical derivation
of general expressions for scale--homogeneous 
operator $P$ and polynomials $P_{\alpha}$ 
according to their inverse length dimensions 
\begin{eqnarray}
[P]=[R], \quad [P_{\alpha}] = [R]-[{\cal G}_{\alpha}] +[u].
\label{dimP}
\end{eqnarray}
Then substituting all the derived
quantities into the deformation 
equation \p{maineq}, extracting equations for   
coefficient--functions of the operator $P$ and polynomials 
$P_{\alpha}$ and, at last, solving them, one can
finally obtain a desirable explicit expression for the recursion operator
$R$ \p{recrel1}.    

Because of the quite technically complicated construction of the recursion
operator we described above, in what follows we actually make our choice
by first imposing conditions on $\{P,~P_{\alpha}\}$, which are more simple
than the deformation equation \p{maineq}, such that the relation 
\p{recrel1} is satisfied only for a set of special symmetries. Then, after
solving these conditions and obtaining the explicit expression for the
recursion operator we prove additionally that it indeed satisfies the
deformation equation \p{maineq} as well, i.e. proving that associated
form--valued vector field is in fact a generalized (with respect to its
form--valuedness) symmetry of equation \p{phi}. We refer the interested
reader to ref. \cite{kersor3} for all details of the complete computations
of the results of the next sections.
	
\section{\bf Nonlocal Hamiltonians and flows of the $N=2$ ${\alpha}=1$ KdV
hierarchy}

In this section we realize the first step of the general scheme,
presented in the previous section, i.e. construct Hamiltonians and
symmetries of the $N=2$ ${\alpha}=1$ KdV equation \p{flows} which are
relevant in the context of the further construction of the
recursion operator of the $N=2$ ${\alpha}=1$ KdV hierarchy.

\subsection{\bf Hamiltonians}

To this aim let us first define a dimension of the recursion operator
as well as dimensions of Hamiltonians.
Remembering that the recursion operator has to connect two nearest
subsequent bosonic flows ${\textstyle{\partial\over\partial t_{2p-1}}}J$ 
and ${\textstyle{\partial\over\partial t_{2p+1}}}J$ \p{bosflows}
of the $N=2$ ${\alpha}=1$ KdV hierarchy and that 
$[{\textstyle{\partial\over\partial t_{2p-1}}}J]=2p$, 
one can easily establish its inverse length dimension $[R] = 2$.
Moreover, because the flows 
${\textstyle{\partial\over\partial t_{2p+1}}}J$ \p{bosflows}
depend on the Grassmann coordinates
$\{\theta, \overline \theta \}$ only implicitly via its dependence on the
superfield $J(Z)$ as well as the fermionic covariant derivatives 
$\{D,{\overline D}\}$, the recursion operator has to possess the same
property as well. Then using inequality \p{uneq} and relation  
\p{dimP} one can evaluate dimensions of Hamiltonians 
\begin{eqnarray}
0, \quad \frac{1}{2}, \quad 1, \quad \frac{3}{2}, \quad 2, 
\quad \frac{5}{2}, \quad 3
\label{range1}
\end{eqnarray}
and the corresponding  dimensions of the polynomials $P_{\alpha}$
\begin{eqnarray}
3, \quad \frac{5}{2},\quad 2, \quad \frac{3}{2}, \quad 1, 
\quad \frac{1}{2}, \quad 0
\label{range2}
\end{eqnarray}
we are interested in. Keeping in mind that  
the polynomial $P_{\alpha}$, which corresponds to 
the Hamiltonian with the highest dimension entering into eq. \p{recrel1},
has to satisfy the symmetry equation \p{symeq}
(see the third remark at the end of the previous section), as well as the 
above--mentioned fact that the recursion operator does not depend
explicitly on $\{\theta, \overline \theta \}$ 
and that the minimal dimension of the 
$\{ {\theta}, {\overline {\theta}} \}$--independent flows is\footnote{We
have verified by explicit construction that there are no flows with 
dimensions $\{0, ~\frac{1}{2}\}$ and that existing flows with 
dimensions $\{1,~\frac{3}{2}\}$ are 
$\{ {\theta}, {\overline {\theta}} \}$--dependent
$\Bigl($see the expressions for the flows 
$U_0J$, $D_{\frac{1}{2}}J$ and ${\overline D}_{\frac{1}{2}}J$ 
($[U_0J]=1,~ [D_{\frac{1}{2}}J]=
[{\overline D}_{\frac{1}{2}}J]=\frac{3}{2}$) in eqs.
(\ref{gnlskdvnfirst}--\ref{fermflows2})
and discussion at the end of the next subsection$\Bigr)$.} 2
(it is the dimension of the first bosonic flow
${\textstyle{\partial\over\partial t_1}}{J}$
\p{flows1}), we are led to the final conclusion that for our ultimate
purposes we have to know a complete set of superfield Hamiltonians with
inverse length dimensions
\begin{eqnarray}
0, \quad \frac{1}{2}, \quad 1
\label{range3}
\end{eqnarray}
only. We would like especially to remark that all superfield components
of these Hamiltonians have in general to be included into the {\it Ansatz} 
\p{recrel1} for the recursion operator.

Superfield Hamiltonians 
\begin{eqnarray}
I_{m} &=& \int^{+\infty}_{-\infty} dz {\cal I}_{m+1}, \quad m \in
\frac{\hbox{\bbd N}}{2}   
\label{seriesint}
\end{eqnarray}
with dimensions $\{0, \frac{1}{2}\}$ were constructed in \cite{ks}.
Their Hamiltonian densities, which are unrestricted $N=2$ superfields
containing four independent superfield components, are
\begin{eqnarray}
{\cal I}_{1} = J, \quad 
{\textstyle{\partial\over\partial t_3}}{J} =
\Bigl(J~'' + 3 J[D,{\overline D}~] J + J^3\Bigr)~'
\label{int0}
\end{eqnarray}
and
\begin{eqnarray}
{\cal I}_{\frac{3}{2}}
= e^{+2{\partial}^{-1}J}DJ, \quad
{\textstyle{\partial\over\partial t_3}}{\cal I}_{\frac{3}{2}} =
\Bigl({\cal I}_{\frac{3}{2}}~'' - 3J{\cal I}_{\frac{3}{2}}~'
+ 3J^2{\cal I}_{\frac{3}{2}}
+3([D,{\overline D}] J){\cal I}_{\frac{3}{2}}\Bigr)~'
\label{int1}
\end{eqnarray}
as well as
\begin{eqnarray}
{\cal I}^{\star}_{\frac{3}{2}} = e^{-2{\partial}^{-1}J}{\overline D}J,
\quad {\textstyle{\partial\over\partial t_3}}
{\cal I}^{\star}_{\frac{3}{2}} =
\Bigl({\cal I}^{\star}_{\frac{3}{2}}~'' 
+ 3J{\cal I}^{\star}_{\frac{3}{2}}~'
+ 3J^2{\cal I}^{\star}_{\frac{3}{2}}
+3([D,{\overline D}] J){\cal I}^{\star}_{\frac{3}{2}}\Bigr)~'.
\label{int2}
\end{eqnarray}
What concerns to the remaining dimension $1$ in \p{range3}, we have  
constructed the corresponding Hamiltonian $I_1$ \p{seriesint} with the
density
\begin{eqnarray}
{\cal I}_{2}&=&
{\cal I}^{\star}_{\frac{3}{2}} {\partial}^{-1}{\cal I}_{\frac{3}{2}}
+({\overline D}J){\partial}^{-1}DJ, \nonumber\\
{\textstyle{\partial\over\partial t_3}}{\cal I}_{2} &=&
\Bigl[2({\overline D}J)DJ~' -2({\overline D}J)~'DJ + 
4J({\overline D}J)DJ\nonumber\\
&-&({\partial}^{-1}{\cal I}_{\frac{3}{2}})
{\partial}^{-1}{\textstyle{\partial\over\partial t_{3}}}
{\cal I}^{\star}_{\frac{3}{2}}-(D{\partial}^{-1}J){\overline D}
{\partial}^{-1}{\textstyle{\partial\over\partial t_{3}}}J\Bigr]~'
\label{int3}
\end{eqnarray}
by "brute-force".
Hereafter, the subscripts denote inverse length dimensions.
The Hamiltonians $I_{\frac{1}{2}}$ and $I^{\star}_{\frac{1}{2}}$
\p{seriesint} with the densities ${\cal I}_{\frac{3}{2}}$ (\ref{int1}) and 
${\cal I}_{\frac{3}{2}}$ (\ref{int2}), respectively, are
related by the complex conjugation \p{conjstar}, while the complex
conjugation properties of the Hamiltonians $I_0,I_1$ are 
$I^{\star}_0 = -I_0$, $I^{\star}_1 = -I_1$. 
We have verified by explicit construction that $I_0$, $I_{\frac{3}{2}}$,
$I^{\star}_{\frac{3}{2}}$ and $I_{1}$ are the only Hamiltonians
with dimensions $\{0, \frac{1}{2}, 1\}$.

To close this subsection 
let us present Fr\'echet derivatives $f_i$ ($i=\frac{1}{2}, 1$) of the
Hamiltonian densities ${\cal I}_{i+1}$ (\ref{int1}--\ref{int3})
\begin{eqnarray}
&&f_{\frac{1}{2}}= e^{2({\partial}^{-1}J)}D+
2{\cal I}_{\frac{3}{2}}{\partial}^{-1}, \quad
f^{\star}_{\frac{1}{2}}= e^{-2({\partial}^{-1}J)} {\overline D}
-2{\cal I}^{\star}_{\frac{3}{2}}{\partial}^{-1}, \nonumber\\
&& f_1= {\cal I}^{\star}_{\frac{3}{2}}{\partial}^{-1}f_{\frac{1}{2}}
-({\partial}^{-1}{\cal I}_{\frac{3}{2}})f^{\star}_{\frac{1}{2}}
+({\overline D}J)D{\partial}^{-1} -(D{\partial}^{-1}J){\overline D}
\label{frechet}
\end{eqnarray}
as well as their operator conjugated quantities 
\begin{eqnarray}
&&f^{T}_{\frac{1}{2}}=-De^{2({\partial}^{-1}J)}-
2{\partial}^{-1}{\cal I}_{\frac{3}{2}}, \quad
{f^{\star}_{\frac{1}{2}}}^{T}=-{\overline D}e^{-2({\partial}^{-1}J)} 
+2{\partial}^{-1}{\cal I}^{\star}_{\frac{3}{2}}, \nonumber\\
&& f^{T}_1= 
f^{T}_{\frac{1}{2}}{\partial}^{-1}{\cal I}^{\star}_{\frac{3}{2}}
+{f^{\star}_{\frac{1}{2}}}^{T}({\partial}^{-1}{\cal I}_{\frac{3}{2}})
-D{\partial}^{-1}({\overline D}J)
-{\overline D}(D{\partial}^{-1}J)
\label{frechetT}
\end{eqnarray}
which we use in what follows.
Here, we use the following standard convention regarding the 
operator conjugation (transposition) $^T$
\begin{eqnarray}
({\partial},D,{\overline D})^{T}=-({\partial},D,{\overline D}),\quad
{(OP)}^{T}=(-1)^{d_{O}d_{P}}P^{T}O^{T}
\label{transp}
\end{eqnarray}
where $O$ ($P$) is an arbitrary
operator with the Grassmann parity $d_{O}$ 
($d_P$), $d_{O}$=0 ($d_{O}=1$)
for bosonic (fermionic) operator ${O}$.

\subsection{\bf Flows}

The flows can be derived from Hamiltonians by means
of the formula \p{hamstr}. Let us present fermionic
and bosonic flows generated by the superfield integrals with the
densities ${\cal I}_{0}$, ${\cal I}_{\frac{1}{2}}$, 
${\cal I}^{\star}_{\frac{1}{2}}$ and ${\cal I}_1$ 
(\ref{int0}--\ref{int3}),  
\begin{eqnarray}
U_0 J = ({\theta}\frac{\partial}{\partial\theta}-
{\overline \theta} \frac{\partial}{\partial\overline\theta})J, \quad
Q_{\frac{1}{2}}J = QJ, \quad 
{\overline Q}_{\frac{1}{2}}J = {\overline Q}J, 
\quad {\textstyle{\partial\over\partial t_1}}{J} =J~'
\label{gnlskdvnfirst}
\end{eqnarray}
and
\begin{eqnarray}
D_{\frac{1}{2}}J = {\cal D}_{\frac{5}{2}} (\theta \overline \theta), \quad 
U^{(+)}_{1}J = {\cal D}_{\frac{5}{2}} (\theta), \quad  
U^{(-)}_{1}J = {\cal D}_{\frac{5}{2}} (\overline \theta), \quad
D_{\frac{3}{2}}J = {\cal D}_{\frac{5}{2}} (1)
\label{fermflows1}
\end{eqnarray}
and
\begin{eqnarray}
{\overline D}_{\frac{1}{2}}J = {\overline {\cal D}}_{\frac{5}{2}} 
(\theta \overline \theta), \quad 
{\overline U}^{(+)}_{1}J = {\overline {\cal D}}_{\frac{5}{2}} 
(\overline \theta),
\quad {\overline U}^{(-)}_{1}J = 
{\overline {\cal D}}_{\frac{5}{2}}(\theta), \quad 
{\overline D}_{\frac{3}{2}}J = 
{\overline {\cal D}}_{\frac{5}{2}}(1)
\label{fermflows2}
\end{eqnarray}
as well as
\begin{eqnarray}
U_1J = {\cal D}_{3} (\theta \overline \theta), \quad 
Q_{\frac{3}{2}}J = {\cal D}_{3} (\overline \theta), \quad  
{\overline Q}_{\frac{3}{2}}J = {\cal D}_{3} (\theta),\quad
U_{2}J = {\cal D}_{3} (1),
\label{newflows}
\end{eqnarray}
respectively, where  $D_{\frac{p}{2}}$ and ${\overline D}_{\frac{p}{2}}$, 
$Q_{\frac{p}{2}}$ and
${\overline Q}_{\frac{p}{2}}$ ($U_p$, $U^{(\pm)}_{p}$ and  
${\overline U}^{(\pm)}_{p}$) ($p \in \hbox{\bbd N}$) are
new fermionic (bosonic) evolution derivatives
with the following properties with respect to the complex conjugation
\p{conjstar}: 
\begin{eqnarray}
D_{\frac{p}{2}}^{\star} = {\overline {\cal D}}_{\frac{p}{2}}, \quad 
Q_{\frac{p}{2}}^{\star} = {\overline {\cal Q}}_{\frac{p}{2}}, \quad  
{U^{\pm}_{p}}^{\star} = {\overline  U}^{\pm}_{p}, \quad 
{U_{p}}^{\star} = (-1)^{p+1} U_{p}, \quad 
{\textstyle{\partial\over\partial t_{2p+1}}}^{\star} =
-{\textstyle{\partial\over\partial t_{2p+1}}}, 
\label{invrel}
\end{eqnarray}
$Q$ and ${\overline Q}$ are generators of
the $N=2$ supersymmetry,
\begin{eqnarray}
&&Q=\frac{\partial}{\partial\theta}
 +\frac{1}{2}\overline\theta\frac{\partial}{\partial z}, \quad
{\overline Q}=\frac{\partial}{\partial\overline\theta}
 +\frac{1}{2}\theta\frac{\partial}{\partial z}, \quad
Q^{2}={\overline Q}^{2}=0, \quad
\left\{ Q,{\overline Q} \right\} = {\partial}, \nonumber\\
&&\left\{ Q,D \right\} =\left\{ Q,{\overline D} \right\} = 0, \quad
\left\{ {\overline Q},D \right\} =
\left\{ {\overline Q},{\overline D}\right\} = 0,
\label{QQ}
\end{eqnarray}
and in eqs. (\ref{gnlskdvnfirst}--\ref{newflows}) we have introduced the
operators 
\begin{eqnarray}
{\cal D}_{\frac{5}{2}} \equiv - J_2 f^{T}_{\frac{1}{2}}, \quad
{\overline {\cal D}}_{\frac{5}{2}} \equiv
J_2 {f^{\star}_{\frac{1}{2}}}^{T}, \quad 
{\cal D}_3 \equiv J_2 f^{T}_{1} 
\label{definition}
\end{eqnarray}
where the operators $J_2$ and $f^{T}_{i}$ are defined in equations 
\p{hamstr2} and \p{frechetT}, respectively. When deriving eqs.  
(\ref{gnlskdvnfirst}--\ref{newflows}) we integrated by parts and made
essential use of the following realization for the inverse derivative:
\begin{eqnarray}
{\partial}^{-1}_{z} \equiv 
\frac{1}{2}\int^{+\infty}_{-\infty} dx~\epsilon(z-x), \quad
\epsilon(z-x) = -\epsilon(x-z)\equiv 1, \quad if \quad z> x.
\label{bosder}
\end{eqnarray}

From the derived formulae (\ref{gnlskdvnfirst}--\ref{newflows}) one can
easily see that only the flows, generated by the highest superfield
components of the Hamiltonians (\ref{int0}--\ref{int3}), depend implicitly
on the Grassmann coordinates $\{\theta, \overline \theta \}$,
while other flows comprise the latter explicitly.
The minimal dimension of the 
$\{ {\theta}, {\overline {\theta}}\}$--independent flows is 2 and it
corresponds to the first bosonic flow
${\textstyle{\partial\over\partial t_1}}{J}$ \p{gnlskdvnfirst}.

\subsection{\bf The algebra of flows and Hamiltonians}

Using the explicit expressions (\ref{gnlskdvnfirst}--\ref{fermflows2}) for
the fermionic flows $Q_{\frac{1}{2}}$, ${\overline Q}_{\frac{1}{2}}$,
$D_{\frac{1}{2}}$ and ${\overline D}_{\frac{1}{2}}$
one can calculate the folowing anticommutators
\begin{eqnarray}
D_{\frac{1}{2}}^2={\overline D}_{\frac{1}{2}}^2= 0, \quad
\{D_{\frac{1}{2}},{\overline D}_{\frac{1}{2}}\}=
+{\textstyle{\partial\over\partial t_1}}
\label{alg1}
\end{eqnarray}
and
\begin{eqnarray}
Q_{\frac{1}{2}}^2={\overline Q}_{\frac{1}{2}}^2= 0, \quad
\{Q_{\frac{1}{2}},{\overline Q}_{\frac{1}{2}}\}=
-{\textstyle{\partial\over\partial t_1}}
\label{alg2}
\end{eqnarray}
and observe that they form two sets of closed algebraic relations,
moreover each of them displays the $N=2$ supersymmetry. 
Thus we are led to the important conclusion that 
the $N=2$ ${\alpha}=1$ KdV hierarchy possesses actually a more rich
symmetry structure, than it is indicated in its title, related
to two different $N=2$ supersymmetries with the generators
$\{Q_{\frac{1}{2}},~{\overline Q}_{\frac{1}{2}}, 
{\textstyle{\partial\over\partial t_1}}\}$ and
$\{D_{\frac{1}{2}},~{\overline D}_{\frac{1}{2}}, 
{\textstyle{\partial\over\partial t_1}}\}$. 
 
The Poisson bracket algebra \p{hamstr2} can also be used to derive the
following useful formula
\begin{eqnarray}
\Bigl\{\int dZ {\widetilde {\cal H}}_1, \int dZ{\widetilde {\cal H}}_2
\Bigr\}=\int dZ ({\widetilde f}_1 J_2 {\widetilde f}^{T}_2)_0
\label{Pbrack}
\end{eqnarray}
where ${\widetilde f}_1$ and ${\widetilde f}^{T}_2$ are  
Fr\'echet derivatives of the Hamiltonian densities 
${\widetilde {\cal H}}_1$ and ${\widetilde {\cal H}}_2$, respectively.
Using this formula one can calculate
the Poisson brackets between the integrals $I_0$, $I_{\frac{1}{2}}$,
$I^{\star}_{\frac{1}{2}}$ and $I_{1}$ (\ref{int0}--\ref{int3}), and this
algebra is isomorphic to the algebra of the corresponding flows.
Repeatedly applying this procedure one can derive new nonlocal
Hamiltonians. As an illustrative example we present 
the Hamiltonian density of the Hamiltonian 
$I_2$ ($I^{\star} = -I_2$)  \p{seriesint} 
\begin{eqnarray}
{\cal I}_{3}&=&
(D{\cal I}^{\star}_{\frac{3}{2}}) 
{\partial}^{-1}{\overline D}{\cal I}_{\frac{3}{2}}
+({\overline D}{\cal I}^{\star}_{\frac{3}{2}}) 
{\partial}^{-1}D{\cal I}_{\frac{3}{2}}
+\frac{1}{2}[D,{\overline D}]J^2
-\frac{3}{2}J[D,{\overline D}]J\nonumber\\
&+&J({\partial}^{-1}DJ){\overline D}J
-J(DJ){\partial}^{-1}{\overline D}J 
-([D,{\overline D}]J)({\partial}^{-1}DJ){\partial}^{-1}{\overline D}J
\label{int4}
\end{eqnarray}
derived in this way. {\it It is interesting to remark that the higher
superfield component of the Hamiltonian $I_2$ ($I_0$) coincides with the
Hamiltonian $H_3$ ($H_1$) \p{bosham03}.}

A discussion of the complete, very rich superalgebraic structure of the
$N=2$ ${\alpha}=1$ KdV hierarchy is out of the scope of the present Letter
and will be discussed in \cite{kersor2}. 

\section{\bf Recursion operator and bi-Hamiltonian structure of the $N=2$
${\alpha}=1$ KdV hierarchy}

Now, the results of preceding sections give us all necessary inputs
to construct the recursion operator of the $N=2$ ${\alpha}=1$ KdV
hierarchy following the algorithm described in Section 3 and then to
derive its bi-Hamiltonian structure.  
In this Section we present main results in a telegraphic style and refer
the reader to ref. \cite{kersor3} for more details.

\subsection{\bf Recursion operator}

We use the {\it Ansatz} \p{recrel1} for the recursion operator $R$ where
only superfield components of the Hamiltonians $I_0$, $I_{\frac{1}{2}}$,
$I^{\star}_{\frac{1}{2}}$ and $I_{1}$ (\ref{int0}--\ref{int3})
are included. Then, as it was already noted at the end of Section 3,  
we impose the condition that two special symmetries
(${\textstyle{\partial\over\partial t_{5}}}J$ and
${\textstyle{\partial\over\partial t_{7}}}J$ \p{bosflows})
of the $N=2$ ${\alpha}=1$ KdV equation \p{flows}
are related by this recursion operator as
\begin{eqnarray}
{\textstyle{\partial\over\partial t_{7}}}J\equiv
res(L^{7}) = R~ res(L^{5})
\equiv R~ {\textstyle{\partial\over\partial t_{5}}}J
\label{eqsolv}
\end{eqnarray}
in accordance with eq. \p{recrel1}, and solve this condition. It is
interesting to remark that this condition completely fixes all unknown
coefficient--functions, involved in the {\it Ansatz} for the recursion
operator, and allows to construct the explicit expression for the latter
\begin{eqnarray}
R &=&{\partial}^2 +2J[D,{\overline D}~]
+ \frac{3}{2}([D,{\overline D}~]J) + J^2\nonumber\\ 
&+& \Bigl\{J~'[D,{\overline D}~] 
+ \frac{1}{2}([D,{\overline D}~]J~')
+ \frac{1}{2}[D,{\overline D}~]J~'
-D ({\overline D}J^2) -{\overline D}(DJ^2) 
\Bigr\}{\partial}^{-1} \nonumber\\&+& 
{\overline D}\Bigr\{{\cal I}^{\star}_{\frac{3}{2}}
[(D{\partial}^{-1}J) - \frac{1}{2}D]
-D{\cal I}^{\star}_{\frac{3}{2}}\Bigl\}{\partial}^{-1}
f_{\frac{1}{2}} + D\Bigl\{ {\cal I}_{\frac{3}{2}}
[({\overline D}{\partial}^{-1}J) + \frac{1}{2}{\overline D}]
+{\overline D}{\cal I}_{\frac{3}{2}}\Bigr\}{\partial}^{-1} 
f^{\star}_{\frac{1}{2}}\nonumber\\
&-& [{\overline D}~ (DJ)+D~({\overline D}J)]
\Bigr\{ {\partial}^{-1}f_{1}
+({\partial}^{-1}{\cal I}_{\frac{3}{2}})
{\partial}^{-1}f^{\star}_{\frac{1}{2}} 
+(D{\partial}^{-1}J){\overline D}{\partial}^{-1}\Bigl\}
\label{recop1}
\end{eqnarray}
where $f_i$ ($i=\frac{1}{2}, 1$) are
the Fr\'echet derivatives defined in eqs. \p{frechet}.
Then we have explicitly verified that the constructed expression
\p{recop1} satisfies the deformation equation \p{maineq}, i.e. it   
indeed gives the proper recursion operator for symmetries of the 
$N=2$ ${\alpha}=1$ KdV hierarchy.

We would like to note that other conditions, similar to the
condition \p{eqsolv}, which relate flows 
${\textstyle{\partial\over\partial t_{1}}}J$ and
${\textstyle{\partial\over\partial t_{3}}}J$ or/and
flows ${\textstyle{\partial\over\partial t_{3}}}J$ and
${\textstyle{\partial\over\partial t_{5}}}J$ do not fix the
recursion operator completely. 

For completeness let us also present the recurrence relations 
\begin{eqnarray}
Y^{a}_{p+1}J &=& R~Y^{a}_pJ    
\label{recop}
\end{eqnarray}
for flows $Y^{a}_{p}J$ of the $N=2$ supersymmetric ${\alpha}=1$
KdV hierarchy in the following useful form:
\begin{eqnarray}
Y^{a}_{p+1}J &=&
\Bigl\{{\partial}^2 +2J[D,{\overline D}~]
+ \frac{3}{2}([D,{\overline D}~]J)+J^2\Bigr\}Y^{a}_{p}J\nonumber\\ 
&+& \Bigl\{J~'[D,{\overline D}~] 
+ \frac{1}{2}([D,{\overline D}~]J~')
+ \frac{1}{2}[D,{\overline D}~]J~'
- D ({\overline D}J^2) - {\overline D}(DJ^2) \Bigr\}{\partial}^{-1} 
Y^{a}_{p}J \nonumber\\&+&(-1)^{d_{Y^{a}}}{\overline D}
\Bigr\{{\cal I}^{\star}_{\frac{3}{2}}
[(D{\partial}^{-1}J) - \frac{1}{2}D]
-D{\cal I}^{\star}_{\frac{3}{2}}\Bigl\}
{\partial}^{-1}Y^{a}_{p}{\cal I}_{\frac{3}{2}} 
\nonumber\\&+& (-1)^{d_{Y^{a}}}D\Bigl\{ {\cal I}_{\frac{3}{2}}
[({\overline D}{\partial}^{-1}J) + \frac{1}{2}{\overline D}]
+{\overline D}{\cal I}_{\frac{3}{2}}\Bigr\}
{\partial}^{-1}Y^{a}_{p}{\cal I}^{\star}_{\frac{3}{2}}\nonumber\\
&-& [{\overline D}(DJ)+D({\overline D}J)]\Bigr\{ 
{\partial}^{-1}Y^{a}_{p}{\cal I}_{2}
+(-1)^{d_{Y^{a}}}({\partial}^{-1}{\cal I}_{\frac{3}{2}})
{\partial}^{-1}Y^{a}_{p}
{\cal I}^{\star}_{\frac{3}{2}}+(D{\partial}^{-1}J)
{\overline D}{\partial}^{-1} Y^{a}_{p}J\Bigl\} ~~~~~
\label{recursion}
\end{eqnarray}
where $Y^{a}_p$ is an evolution derivative 
from the set \p{invrel} and $d_{Y^{a}}$ is its Grassmann parity.

\subsection{\bf Bi-Hamiltonian structure}

We have observed that the constructed recursion operator 
$R$ \p{recop1} can be represented in the factorized form
\begin{eqnarray}
R=J_2J^{-1}_0, \quad 
J^{-1}_0 = 
[D,{\overline D}~]{\partial}^{-1} + {\partial}^{-1} J_2 {\partial}^{-1}
+\frac{1}{2}{f^{\star}_{\frac{1}{2}}}^{T}{\partial}^{-1}f_{\frac{1}{2}}
-\frac{1}{2}{f^{T}_{\frac{1}{2}}}{\partial}^{-1}f^{\star}_{\frac{1}{2}}
\label{firstHSTR}
\end{eqnarray}
where $J_2$ is the second Hamiltonian structure \p{hamstr}, \p{hamstr2} 
and  $f^{T}_{\frac{1}{2}}$ (${f^{\star}_{\frac{1}{2}}}^{T}$) 
is the operator conjugated Fr\'echet derivative \p{frechetT}.
Then $J^{-1}_0$ ($J_0J^{-1}_0=J^{-1}_0J_0=1$) 
can obviously be treated as the inverse operator of the zero
Hamiltonian structure,
\begin{eqnarray}
&&\{J(Z_1),J(Z_2)\}_0 =J_0(Z_1)
\delta^{N=2}(Z_1-Z_2),\quad 
J^{-1}_0{\textstyle{\partial\over\partial t_{2p-1}}}J =
\frac{\delta}{\delta J} H_{2p+1}.
\label{hamstr0}
\end{eqnarray}
Therefore, we come to the conclusion that the $N=2$ supersymmetric 
${\alpha}=1$ KdV hierarchy is a bi-Hamiltonian system. 
Acting $k$-times with the recursion operator \p{recop1} on the second
Hamiltonian structure $J_2$ \p{hamstr}, \p{hamstr2} of the $N=2$  
${\alpha}=1$ KdV hierarchy, one can derive its $2(k+1)$-th Hamiltonian
structure,
\begin{eqnarray}
&&J_{2(k+1)} = R^k J_2, \quad 
{\textstyle{\partial\over\partial t_{2p+1}}}J = 
\{J, H_{2(p-k)+1}\}_{2(k+1)} \equiv
J_{2(k+1)} \frac{\delta}{\delta J} H_{2(p-k)+1}, \nonumber\\
&& \quad \quad \quad \quad
\{J(Z_1),J(Z_2)\}_{2(k+1)}  =J_{2(k+1)}(Z_1) \delta^{N=2}(Z_1-Z_2).
\label{hamstrn}
\end{eqnarray}

\section{\bf Summary} 

In this Letter we have adapted the general algorithm of
constructing recursion operators to the case of the $N=2$
supersymmetric ${\alpha}=1$ KdV equation in $N=2$ superspace.
Then we have constructed all basic objects which are relevant
to this aim: nonpolynomial and nonlocal, bosonic and fermionic
Hamiltonians (\ref{int0}--\ref{int3}) and symmetries
(\ref{gnlskdvnfirst}--\ref{newflows}). Furthermore we have observed
that the $N=2$ ${\alpha}=1$ KdV hierarchy possesses a more rich
symmetry structure, than it is indicated in its title, related
to two subalgebras of the $N=2$ supersymmetry (\ref{alg1}--\ref{alg2})
of its algebra of symmetries.
Finally we have constructed its recursion operator \p{recop1},
recursion relations \p{recursion} as well as zero Hamiltonian structure 
(\ref{firstHSTR}--\ref{hamstr0}) which were unsolved
longstanding problems.

{}~

\noindent{\bf Acknowledgments.}
We would like to thank I.S. Krasil'shchik for useful discussions. 
A.S. is grateful to University of Twente for the hospitality extended to
him during this research. This work was partially supported by the grants
NWO NB 61-491, FOM MF 00/39, RFBR 99-02-18417, RFBR-CNRS 98-02-22034, PICS
Project No. 593, Nato Grant No. PST.CLG 974874 and the Heisenberg-Landau
program.

\end{document}